\documentclass[11pt,twoside]{article}
\usepackage{asp2010}

\resetcounters

\allowtitlefootnote

\begin{document}

\title{PUCHEROS Early Science: A New Be+sdO Candidate\footnotemark}
\author{Th.~Rivinius,$^{1}$ L.~Vanzi,$^{2}$ J.~Chacon,$^{2}$ P.~Leyton,$^{3}$
  K.~G.~Helminiak,$^{3}$ M.~Baffico,$^{3}$ S.~\v{S}tefl,$^1$ D.~Baade,$^{4}$
  G.~Avila,$^{4}$ and C.~Guirao,$^{4}$
%
%
\affil{$^1$ESO---European Organisation for Astronomical Research in the
  Southern Hemisphere, Santiago, Chile}
\affil{$^2$Department of Electrical Engineering and Center of Astro Engineering, Pontificia Universidad Catolica de Chile}
\affil{$^3$Department of Astronomy and Astrophysics, Pontificia Universidad
  Catolica de Chile}
\affil{$^4$ESO---European Organisation for Astronomical Research in the
Southern Hemisphere, Garching b.\ M\"unchen, Germany}}
\titlefootnote{Partly based on observations collected at the European
  Organisation for Astronomical Research in the Southern Hemisphere, Chile
  under Prog-ID 076.C-0164.}
\begin{abstract}We report on the first scientific results with the recently 
commissioned PUCHEROS spectrograph, mounted at the 50\,cm telescope of the
Pontificia Universidad Catolica near Santiago, Chile. A hitherto unknown
candidate Be+sdO binary was identified, $o$\,Pup. If confirmed, it would be
the fourth member of this class. Such stars have obtained their rapid rotation
through binary mass transfer and now consist of a Be star and a hot subdwarf.
\end{abstract}

\section{Introduction}

PUCHEROS is the newly built {\it Pontificia Universidad Catolica High Echelle
  Resolution Optical Spectrograph} \citep{Vanzietal_inpress}. Most components
  are commercial off-the-shelf, enabling a cost-effective design
  ($\sim$25\,000\,USD for hardware). Light is fed into the instrument by an
  optical fiber, attached to the 50\,cm telescope (the former ESO 50\,cm at La
  Silla) at Santa Martina observatory near Santiago de Chile. See
  Table~\ref{tab:PUCH} for an overview of the instrument.

\begin{table}[t]
\caption{\label{tab:PUCH}Instrumental key values of
  PUCHEROS}{\smallskip}
\begin{center}\small
\begin{tabular}{lc}
\tableline\noalign{\smallskip}
Spectral Resolution $\lambda/\Delta\lambda$ & 20\,000       \\
Spectral coverage (single shot)             & 390--730\,nm  \\
Orders                                      & 60--106        \\
Aperture on sky (diameter)                  & 3.5\arcsec    \\
Limiting mag. (1 hr, $S/N$ = 30)            & $V=9$\,mag      \\
Detector                                    & FLI--PL1001E    \\
total effciency                             &  5\%      \\
\noalign{\smallskip}\tableline
\end{tabular}
\end{center}

\caption{\label{tab:omi}Observations of $o$\,Pup}\smallskip
\begin{center}\small
\begin{tabular}{lrrr}
\tableline\noalign{\smallskip}
\multicolumn{1}{c}{Date} & \multicolumn{1}{c}{MJD} &
\multicolumn{1}{c}{H$\alpha$ EW} & \multicolumn{1}{c}{$V/R$}\\
& & \multicolumn{1}{c}{[\AA]} & \multicolumn{1}{c}{He\,{\sc i}\,6678}\\
\noalign{\smallskip}\tableline\noalign{\smallskip}
2011--3--07 & 55\,627.15 &  $-10.6$ &  1.04 \\
2011--3--29 & 55\,649.01 &  $-11.0$ &  0.95 \\
2011--3--30 & 55\,650.07 &  $-11.3$ &  0.98 \\
2011--4--04 & 55\,655.03 &  $-11.1$ &  1.02 \\
2011--4--07 & 55\,658.05 &  $-9.52$ &  1.05 \\
2011--5--05 & 55\,686.03 &  $-10.2$ &  0.95 \\
2011--6--17 & 55\,729.95 &  $-11.7$ &  1.00 \\
2012--1--20 & 55\,946.09 &  $-11.8$ &  1.02 \\
2012--1--24 & 55\,950.11 &  $-10.5$ &  0.95 \\
\noalign{\smallskip}\tableline
\end{tabular}
\end{center}
\end{table}

One of the {science drivers} for PUCHEROS are {Be stars} \citep[see][for a
  review]{2003PASP..115.1153P}, with the two main purposes of monitoring
objects for activity to follow up, as well as to complement data from larger
telescopes by providing a spectroscopic history to put the data in the context
of temporal behaviour.

\section{Observations}
Observations were carried out from Santa Martina in {commissioning} mode in
the {first half of 2011}. In the second half of the year, during telescope
maintenance and re-coating, the instrument was worked over according to the
commissioning results, and is back in {operation since December 2011}. The
commissioning targets already included several Be stars, which were identified
in archival data as being interesting for long term monitoring.

PUCHEROS has started to monitor eleven bright Be stars by now. While most of
the targets, Be and non-Be stars, are meant to built-up a long-term database,
a few targets were chosen for immediate science return. For Be stars this was
$o$\,Pup. The {observation table} for this target with PUCHEROS is given in
Table~\ref{tab:omi}. 

\section{Analysis}

\begin{figure}[t]
\begin{center}
\includegraphics[viewport=0 144 413 442,clip,angle=0,width=4.5in]{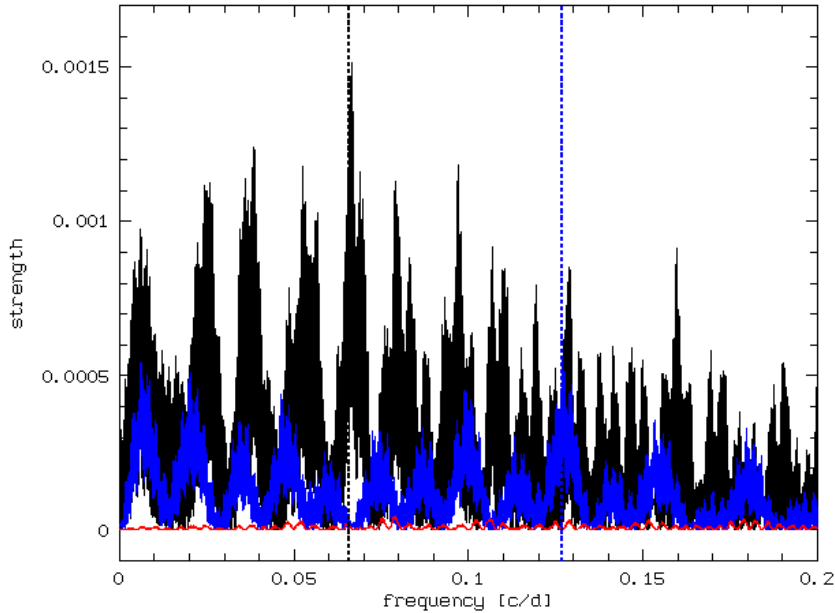}%
\end{center}
\caption{\label{fig:periodo}Periodogram of the $V/R$ values. The original
  (black) and first (blue) and second (red) iteration of period cleaning are
  shown.}
\end{figure}

Based on the appearance of one archival FEROS spectrum, $o$\,Pup was suspected
to be a Be+sdO type binary, of which only three confirmed \citep[$\phi$\,Per,
  59\,Cyg, and
  FY\,CMa:][]{1998ApJ...493..440G,2004A&A...427..307R,2005PAICz..93...21M,2008ApJ...686.1280P}
and one candidate \citep*{1991A&A...250..437W} are known. In such a system the
hot secondary {illuminates and excites a sector of the outer disk edge}. This
causes an additonal emission bump to wander between red and blue with the
orbital period, causing violet-to-red peak height variation ($V/R$)
\citep*{2000A&A...358..208S}. To analyse this, the $V/R$ values, complemented
by the FEROS archival data and amateur measurements available at BeSS
\citep{2011AJ....142..149N}, were subjected to a period analysis (see
Fig.~\ref{fig:periodo}).

The data shows two significant periods, $15.183\pm.013$ and $7.882\pm.002$\,d,
before the residuals are effectively flat. Due to inclusion of data from
seasons before 2011/12, the formal errors are small, but aliasing vs.\ nearby
($\Delta f \approx n\times 1/365$\,d) peaks cannot be entirely excluded. For
this reason, it is actually possible that the shorter period is the first
harmonic of the second. 

Although Fig.~\ref{fig:periodo} looks striking, it was not possibe to fully
exclude the main period to have another value. In fact, only slightly less
significant values for the true period are 26 and 40 days. More data,
currently being taken, might change the relative significance between these
three.

\begin{figure}[t]
\begin{center}
\includegraphics[viewport=80 120 630 680,clip,angle=0,width=2.5in]{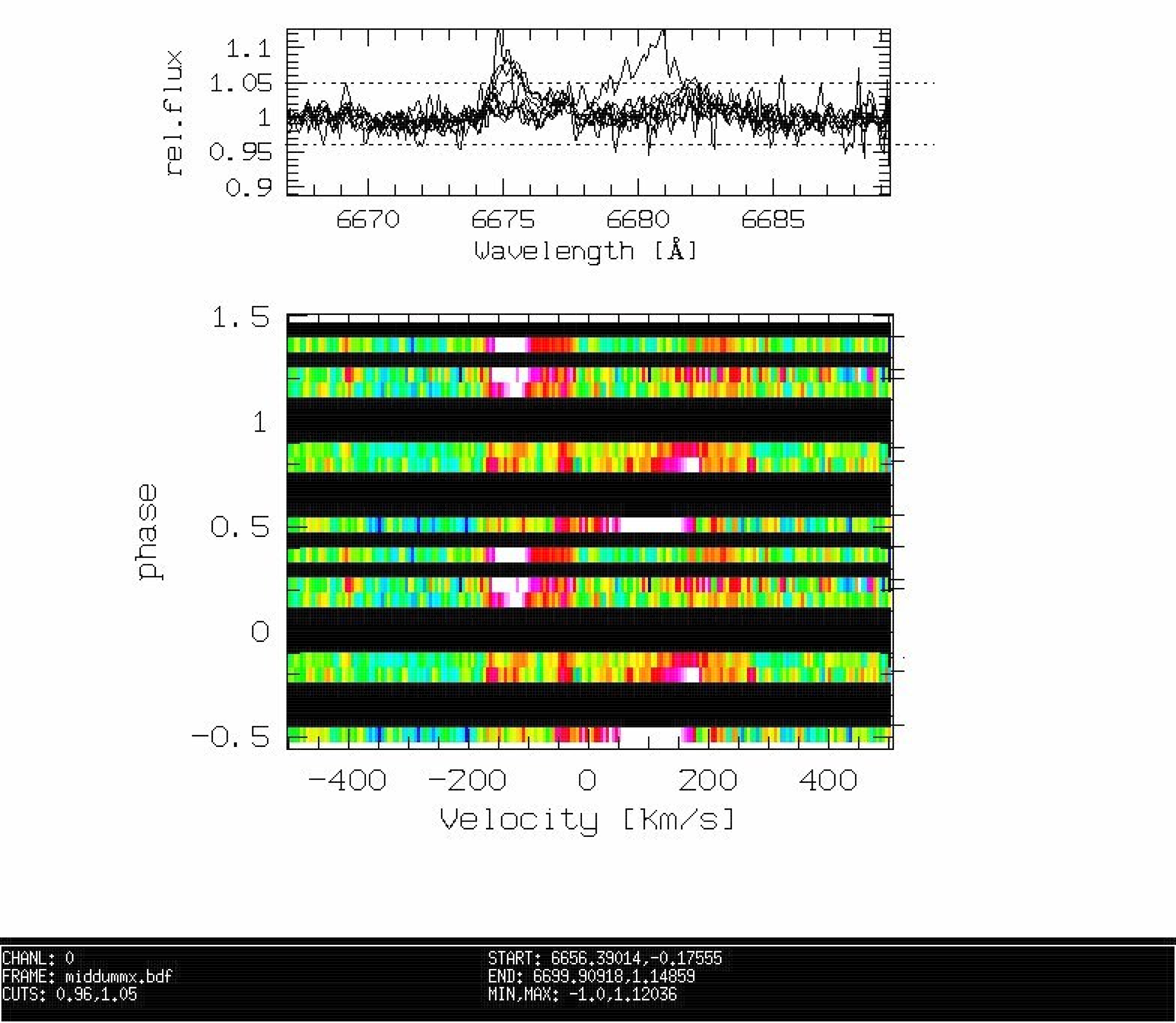}%
\includegraphics[viewport=80 120 630 680,clip,angle=0,width=2.5in]{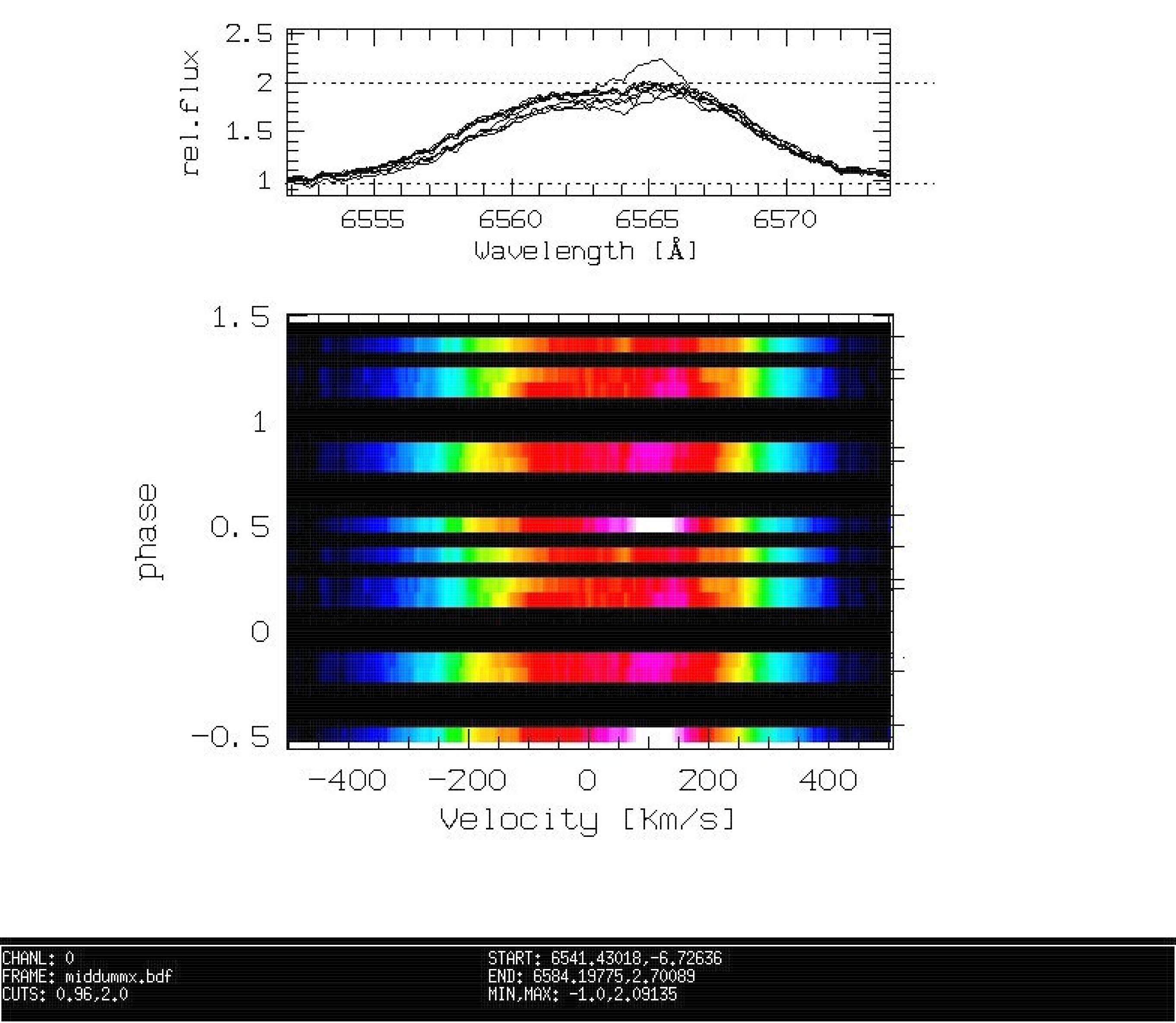}
\end{center}
\caption{\label{fig:dynspec}Phased spectra of He\,{\sc i}\,6678 ({\it left})
  and H$\alpha$ ({\it right}) of $o$\,Pup}
\end{figure}

\section{Results}
The measured $V/R$ ratio in the He\,{\sc i}\,6678 line was subjected to a
period analysis. The strongest peak is at a {period of 15.183\,d}, with a
strong first harmonic. The current data is somewhat ambiguous, as other
periods are possible as well due to strong aliasing (26 and 40\,d). However,
the presence of a {strong and fast periodic $V/R$-variation} has been {firmly
  established}. The possible periods rule out any but a {binary}
hypothesis. Figure~\ref{fig:dynspec} shows the variability of the
\ion{He}{i}\,6678 and H$\alpha$ line profiles, folded with the 15.183\,d
period.

\begin{itemize}
\item In addition to the confirmed systems $\phi$\,Per, 59\,Cyg, and FY\,CMa,
  and the candidate HR\,2142, $o$\,Pup is the {fifth potential Be+sdO binary}.
\item The available data does not allow to distinghuish between three possible
  periods of 15, 26, and 40\,d. Additional observations are currently taken.
\end{itemize}

\acknowledgements The PUCHEROS project was entirely supported by CONICYT
through the Fondecyt project n.~1095187. LV acknowledges support from the
project CONICYT Anillo--ACT 86, JC from the project CONICYT ALMA n.~31090033,
KH acknowledges the Fondecyt postdoctoral grant n.~3120153.

\bibliography{csdyn}

\newpage

\noindent{\bf Note added after submission}\medskip

\noindent
Only on September 23 did we learn about the paper by Koubsk\'y et al.\ (2012,
A\&A 545, A121), which, too, suggests $o$\,Pup to be the fourth Be+sdO
candidate. Because the bulk of their observations, namely the high-cadence
part, was obtained only after our poster was shown during the conference for
which this contribution is submitted, held February 27\,--\,March 2, 2012, we
could not reference their work, which we would otherwise have done as a matter
of course.
\bigskip

Thomas Rivinius
(on behalf of the authors)

\end{document}